\documentclass[]{article}
\usepackage{My-SetUp}
\usepackage[colorlinks=true,allcolors={blue}]{hyperref}

\begin{document}

\title{Principal Eigenvalue and Landscape Function of the Anderson Model on a Large Box}
\author{Daniel S\'anchez-Mendoza}
%\email{dsanchezmendoza@unistra.fr}
%\affiliation{Université de Strasbourg, Institut de Recherche Mathématique Avancée UMR 7501, F-67000 Strasbourg, France.}
\date{}
%\keywords{Anderson model, Integrated density of states, Bernoulli distribution.}

\maketitle

\begin{abstract}
We state a precise formulation of a conjecture concerning the product of the principal eigenvalue and the sup-norm of the landscape function of the Anderson model restricted to a large box. We first provide the asymptotic of the principal eigenvalue as the size of the box grows and then use it to give a partial proof of the conjecture. We give a complete proof for the one dimensional case.	
\end{abstract}

\section{Introduction and Results}
The landscape function, introduced by Filoche and Mayboroda in \cite{Mayboroda1}, has been conjectured to capture the low eigenvalues of the Anderson model operator, discrete or continuous, restricted to a finite large box. We can find this conjecture loosely stated in \cite[Equation 1.4]{Mayboroda2} as: If 0 is the minimum of the support of the potential distribution then
\begin{equation*}
    \lambda_i L_i\approx1+\frac{d}{4} \qquad 1\leq i\ll n^d
\end{equation*}
where $\{\lambda_i\}_{i}$ are the eigenvalues ordered increasingly, $\{L_i\}_{i}$ are the local maxima of the landscape function ordered decreasingly, $d$ is the dimension, and $n$ is the linear size on the box. Numerical experiments with Bernoulli and Uniform potential distributions support the conjecture (see \cite{Mayboroda3},\cite{Mayboroda4}), but to this moment there is no mathematical proof. In this article we give a precise formulation of the conjecture on the discrete setting for the case $i=1$, that is, for the product of the principal (smallest) eigenvalue and the sup-norm of the Landscape function on a large box. We claim such product converges almost surely to an explicit dimensional constant, different from $1+\frac{d}{4}$, as the size of the box goes to infinity and give the proof of the $\liminf$. For a special case in $d=1$, we also give the proof of the $\limsup$.

We start with some definitions and notation. Given a non-empty and finite $A\subseteq\Z^d$ and a positive potential $W:A\rightarrow[0,\infty)$ we consider the Schrödinger operator
\begin{align*}
-\Delta_{A}+W:\ell^2(A)&\longrightarrow\ell^2(A),\\
\phi&\longmapsto (-\Delta_{A}+W)\phi(x)\coloneqq\sum_{ \abs{y-x}=1}\left[\phi(x)-\phi(y)\right]+W(x)\phi(x),
\end{align*}
where $-\Delta_{A}$ has Dirichlet boundary conditions. From it, we define its principal eigenvalue and landscape function
\begin{equation*}
    \lambda_{A,W}\coloneqq\inf\sigma(-\Delta_{A}+W),\qquad L_{A,W}\coloneqq (-\Delta_{A}+W)^{-1}\1_{A}.
\end{equation*}
Notice that $\lambda_{A,W}>0$ and $L_{A,W}$ is always well defined on $A$ since $-\Delta_{A}>0$ and $W\geq0$.

Let $V=\{V(x)\}_{x\in\Z^d}$ be an i.i.d. random non-negative potential whose probability measure and expectation we denote $\mathbb{P}$ and $\mathbb{E}$, and define for $n\in\N$ the box $\Lambda_n\coloneqq[-n,n]^d\cap\Z^d$. Our main objectives are the asymptotics of $ \lambda_{\Lambda_n,V}$ and $\norm{L_{\Lambda_n,V}}_\infty$ as $n\to\infty$, where, as customary, the restriction of $V$ to $\Lambda_n$ is implicit.

In addition to $V$ being non-negative (i.e., $\PPrb{V(0)\in(-\infty,0)}=0$) we will always assume the distribution function $F(t)=\PPrb{V(0)\leq t}$ satisfies one of the following mutually exclusive conditions:
\begin{enumerate}[label=\textbf{(C\arabic*)}]
\item $0<F(0)<1$,\quad(Example: Bernoulli$(p)$ distribution)
\item $F(t)=c\,t^\eta(1+o(1))$ as $t\downarrow 0$ for some $c,\,\eta>0$.\quad(Example: Uniform$(0,1)$ distribution)
\end{enumerate}

We write $n$ instead of $\Lambda_n$ whenever convenient (e.g. $-\Delta_n=-\Delta_{\Lambda_n}$, $\lambda_{n,V}=\lambda_{\Lambda_n,V}$). We denote by $\omega_d$ and $\mu_d$ respectively, the volume of the unit ball in $\R^d$ and the principal eigenvalue of the continuous Laplacian ($-\sum_{i=1}^d\partial^2/\partial x_i^2$) on such ball with Dirichlet boundary conditions.

We now state our conjecture and results. We are always assuming that $V$ is non-negative and satisfies \textbf{(C1)} or \textbf{(C2)}. We claim that:
\begin{Conjecture}\label{Co}
$\displaystyle\lim_{n\to\infty}\lambda_{n,V}\norm{L_{n,V}}_\infty=\frac{\mu_d}{2d}$\quad$\mathbb{P}$-a.s.
\end{Conjecture}
The heuristic argument behind this conjecture is that both $\lambda_{n,V}$ and $\norm{L_{n,V}}_\infty$ are controlled by the largest ball inside of $\Lambda_n$ with zero or very low potential. If the radius of such ball is $r$ then, roughly, $\lambda_{n,V}$ is proportional to $r^{-2}$ and $\norm{L_{n,V}}_\infty$ is proportional to $r^{2}$, making the product of order one in $r$. The appearance of the continuous constant $\frac{\mu_d}{2d}$ is another instance of the solution of a discrete problem converging to the solution of the corresponding continuous one. 
The disagreement between the dimensional constants $\frac{\mu_d}{2d}$ and $1+\frac{d}{4}$ is simply explained by the fact that $1+\frac{d}{4}$ was "guessed" from the numerical experiments, and the two constants are close to each other. For example, for $d=1$ we have $1+\frac{1}{4}=1.25$ and $\frac{\mu_1}{2}=\frac{\pi^2}{8}\approx1.23$.

Using the Min-Max Principle and our hypothesis on $V$ it is straightforward to show that $\lambda_{n,V}$ is decreasing in $n$ and converges to $0$. Our first result is on the speed of this convergence, depending on whether $V$ satisfies \textbf{(C1)} or \textbf{(C2)}:
\begin{Theorem}\label{Th1}\leavevmode
\begin{itemize}
    \item For \textbf{(C1)}, $\displaystyle\lim_{n\to\infty}\lambda_{n,V}\left(\frac{\omega_d\abs{\ln F(0)}}{d\ln n}\right)^{-2/d}=\mu_d$\quad$\mathbb{P}$-a.s.
    \item For \textbf{(C2)}, $\displaystyle\lim_{n\to\infty}\lambda_{n,V}\left(\frac{2\eta \omega_d\ln\ln n}{d^2\ln n}\right)^{-2/d}=\mu_d$\quad$\mathbb{P}$-a.s.
\end{itemize}
\end{Theorem}
The proof of Theorem \ref{Th1} is given in Section 2, and it is divided into the upper and lower bounds of $\lambda_{n,V}$. The upper bound follows form the Min-Max Principle and the previously mentioned heuristic of the largest ball with zero or very low potential. The lower bound is a bit more involved; it uses a Lifshitz tails result form \cite{Biskup} and the connection between the integrated density of states of the (infinite) Anderson model and the distribution function of $\lambda_{n,V}$. 

\begin{figure}[ht]
\centering
\begin{subfigure}{.45\textwidth}
    \centering
    \includegraphics[width=\linewidth]{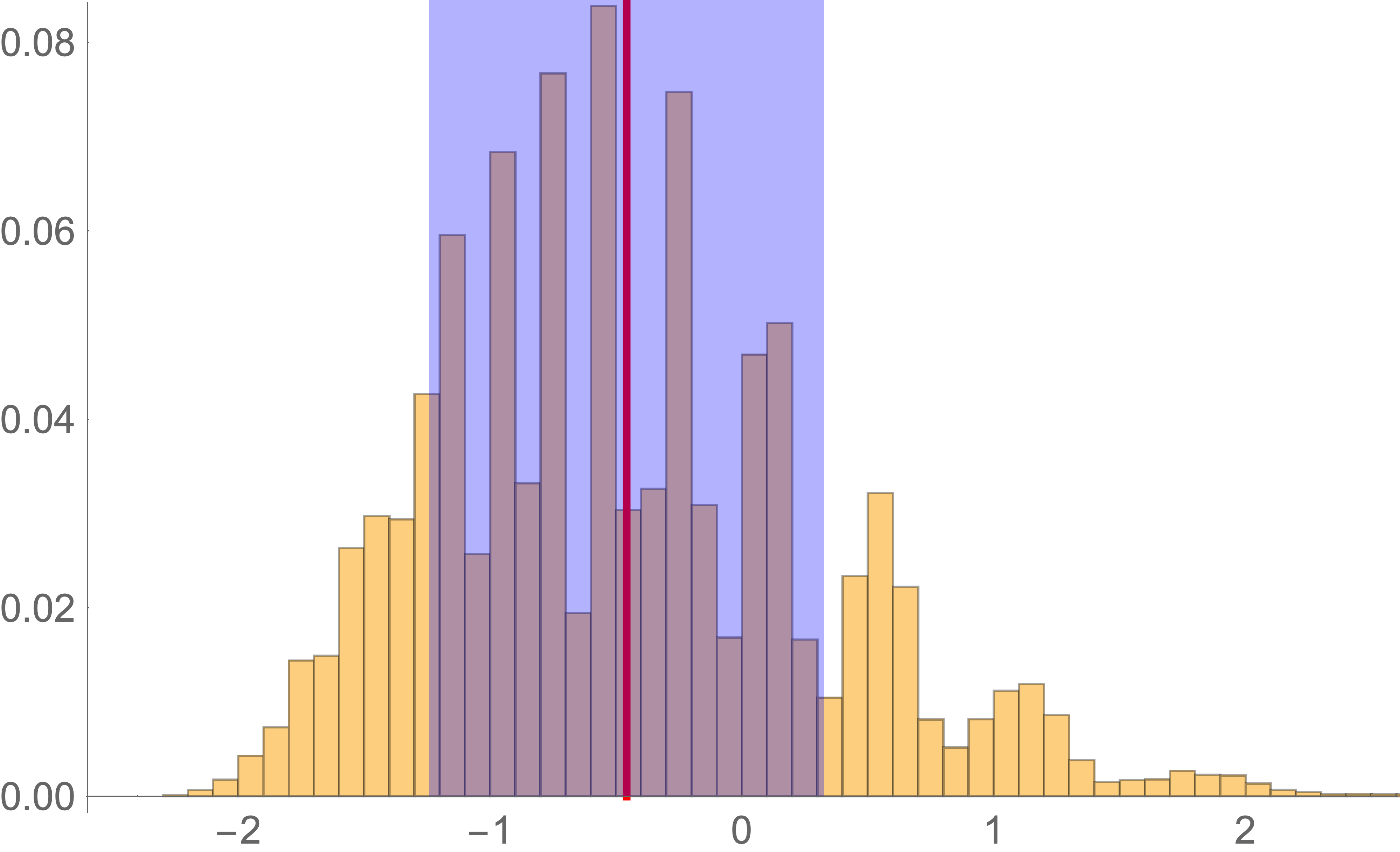}  
    \caption{$n=10^2$, $m=-0.4586$, $s=0.7844$}
\end{subfigure}
\hfill
\begin{subfigure}{.45\textwidth}
    \centering
    \includegraphics[width=\linewidth]{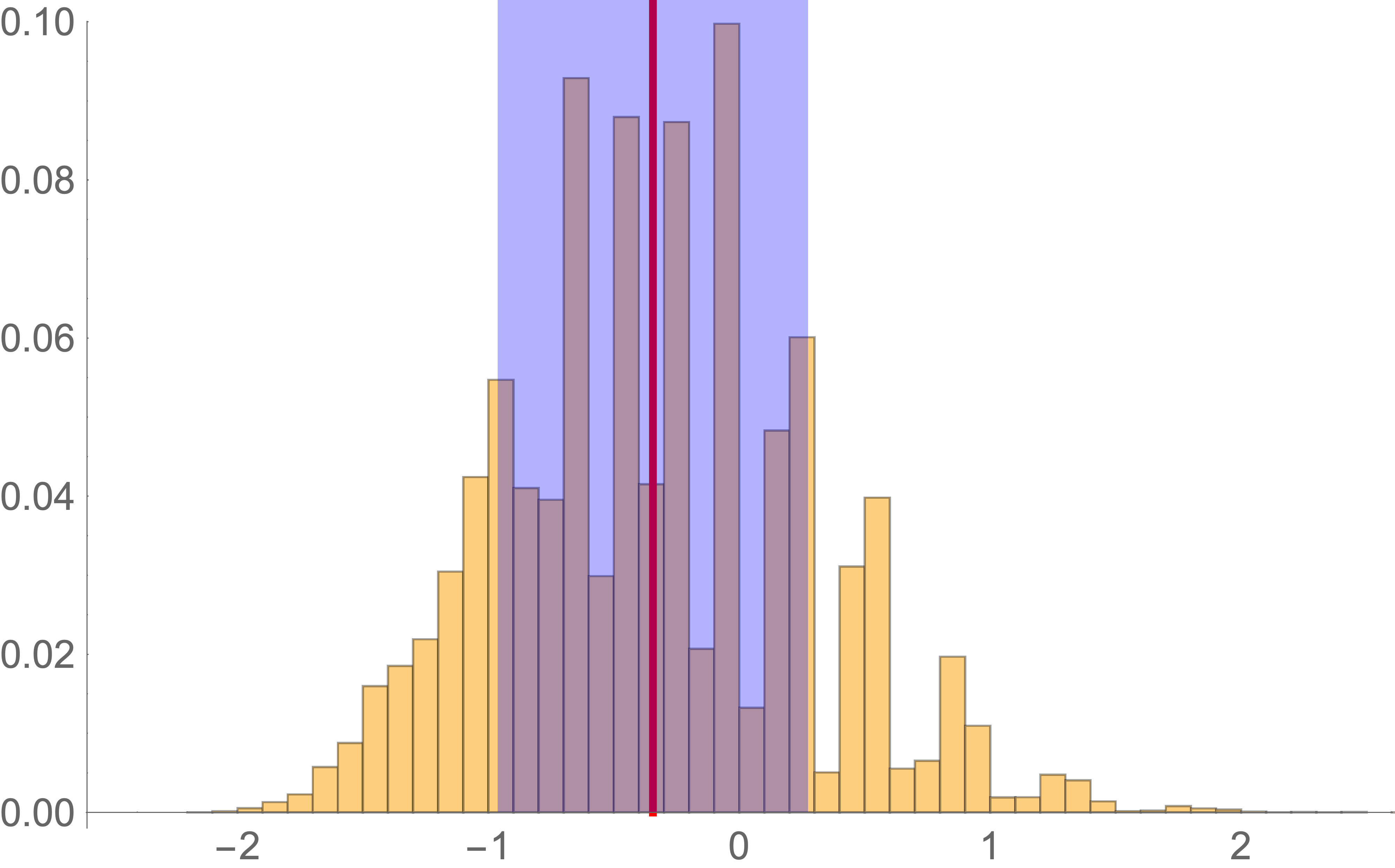}  
    \caption{$n=10^3$, $m=-0.3449$, $s=0.6175$}
\end{subfigure}
\begin{subfigure}{.45\textwidth}
    \centering
    \includegraphics[width=\linewidth]{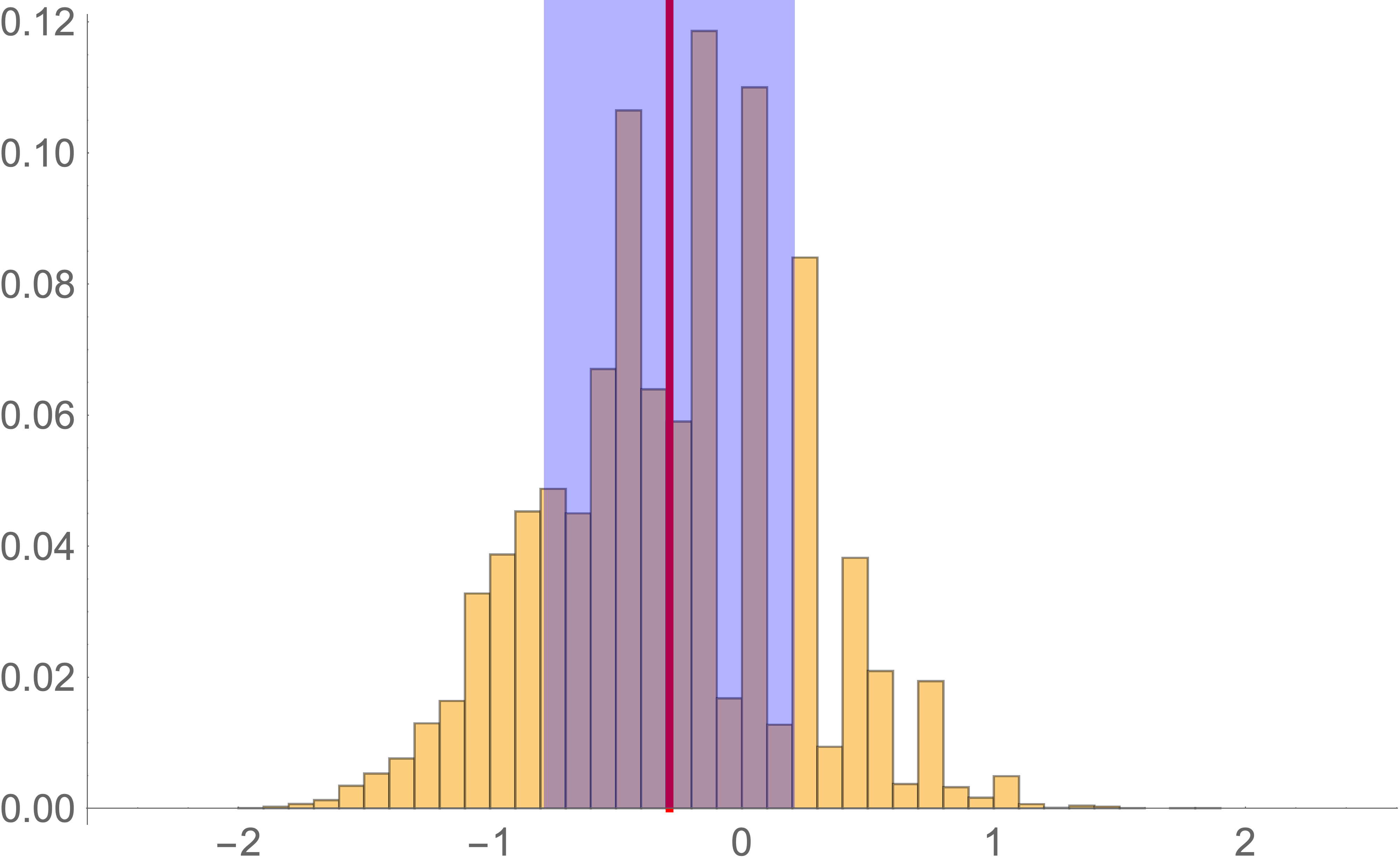}  
    \caption{$n=10^4$, $m=-0.2881$, $s=0.4976$}
\end{subfigure}
\hfill
\begin{subfigure}{.45\textwidth}
    \centering
    \includegraphics[width=\linewidth]{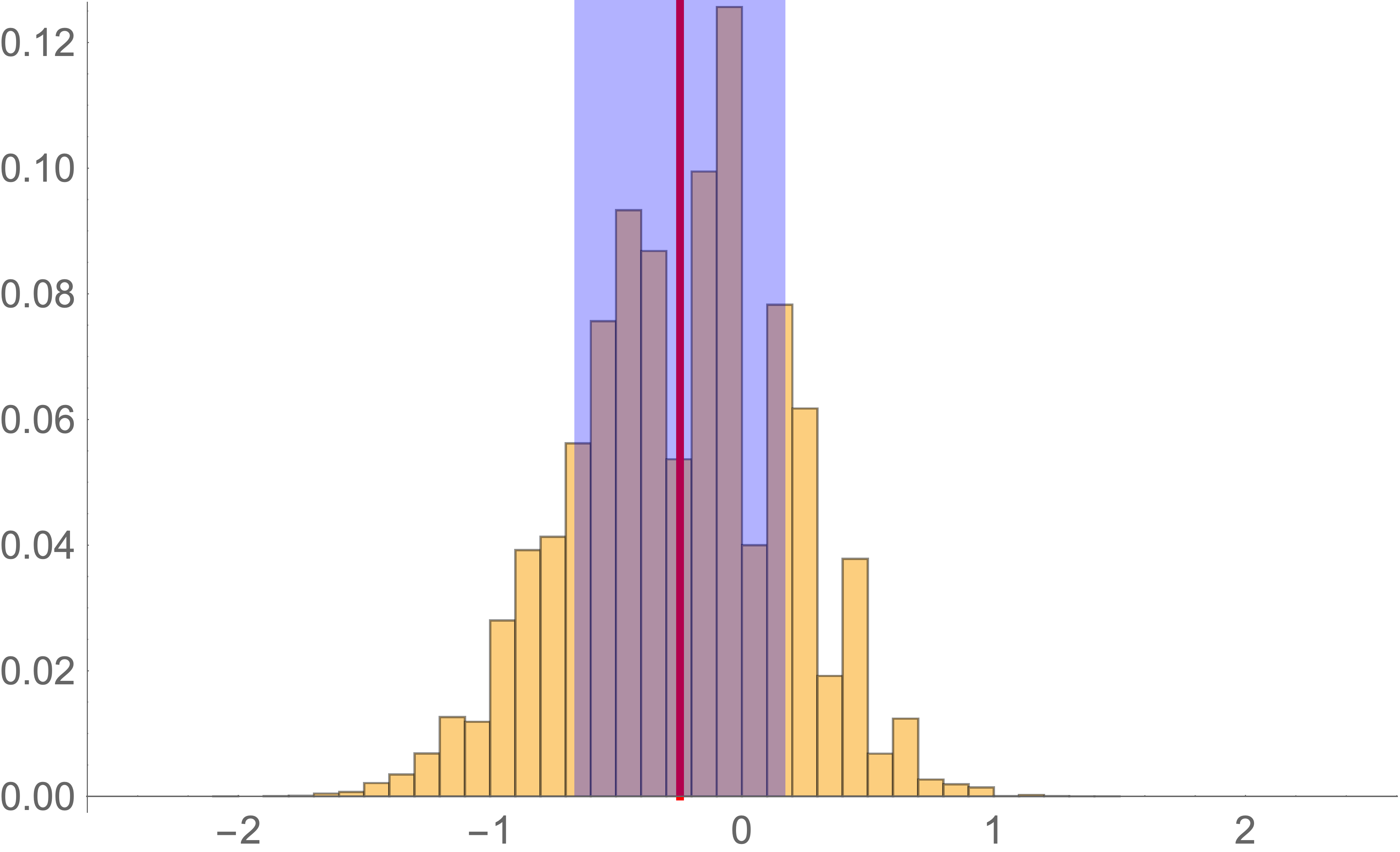}  
    \caption{$n=10^5$, $m=-0.2465$, $s=0.4186$}
\end{subfigure}
\caption{Empirical distribution of $\lambda_{n,V}\left(\frac{\omega_1\abs{\ln F(0)}}{\ln n}\right)^{-2}-\mu_1$ for $d=1$ and $V(0)\overset{\text{d}}{=}$ Bernoulli$(0.3)$ computed form $10^5$ samples. The empirical mean ($m$) and empirical standard deviation ($s$) are shown in red and blue respectively.}
\label{Fig1}
\end{figure}

We tried to illustrate Theorem \ref{Th1}, say in the case $d=1$ and $V(0)\overset{\text{d}}{=}$ Bernoulli$(p)$, by plotting $\lambda_{n,V}\left(\frac{\omega_1\abs{\ln (1-p)}}{\ln n}\right)^{-2}\,\text{v.s.}\,n$ for a single realization of the potential. However, the plot does not show any kind of accumulation up to $n=10^5$, suggesting the convergence is very slow. Instead, we draw the empirical distribution of $\lambda_{n,V}\left(\frac{\omega_1\abs{\ln (1-p)}}{\ln n}\right)^{-2}-\mu_1$ from $10^5$ realizations for $n=10^2,10^3,10^4,10^5$. These are given in Figure \ref{Fig1}, from which we can see that the empirical mean, variance, and distribution concentrate towards $0$ as $n$ increases.

\begin{figure}[ht]
\centering
\begin{subfigure}{.45\textwidth}
    \centering
    \includegraphics[width=\linewidth]{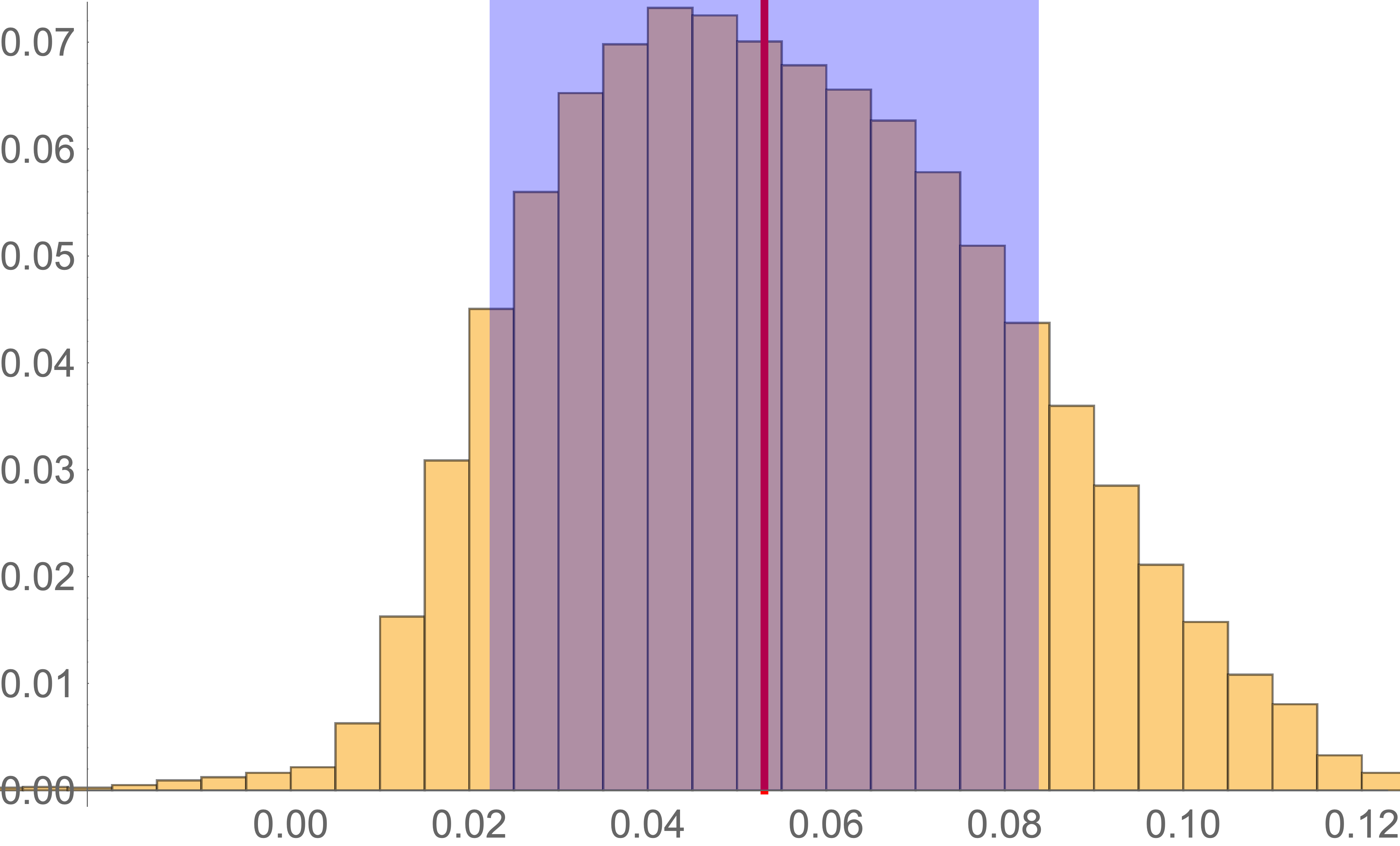}  
    \caption{$n=10^2$, $m=5.3\times10^{-2}$, $s=3.0\times10^{-2}$}
\end{subfigure}
\hfill
\begin{subfigure}{.45\textwidth}
    \centering
    \includegraphics[width=\linewidth]{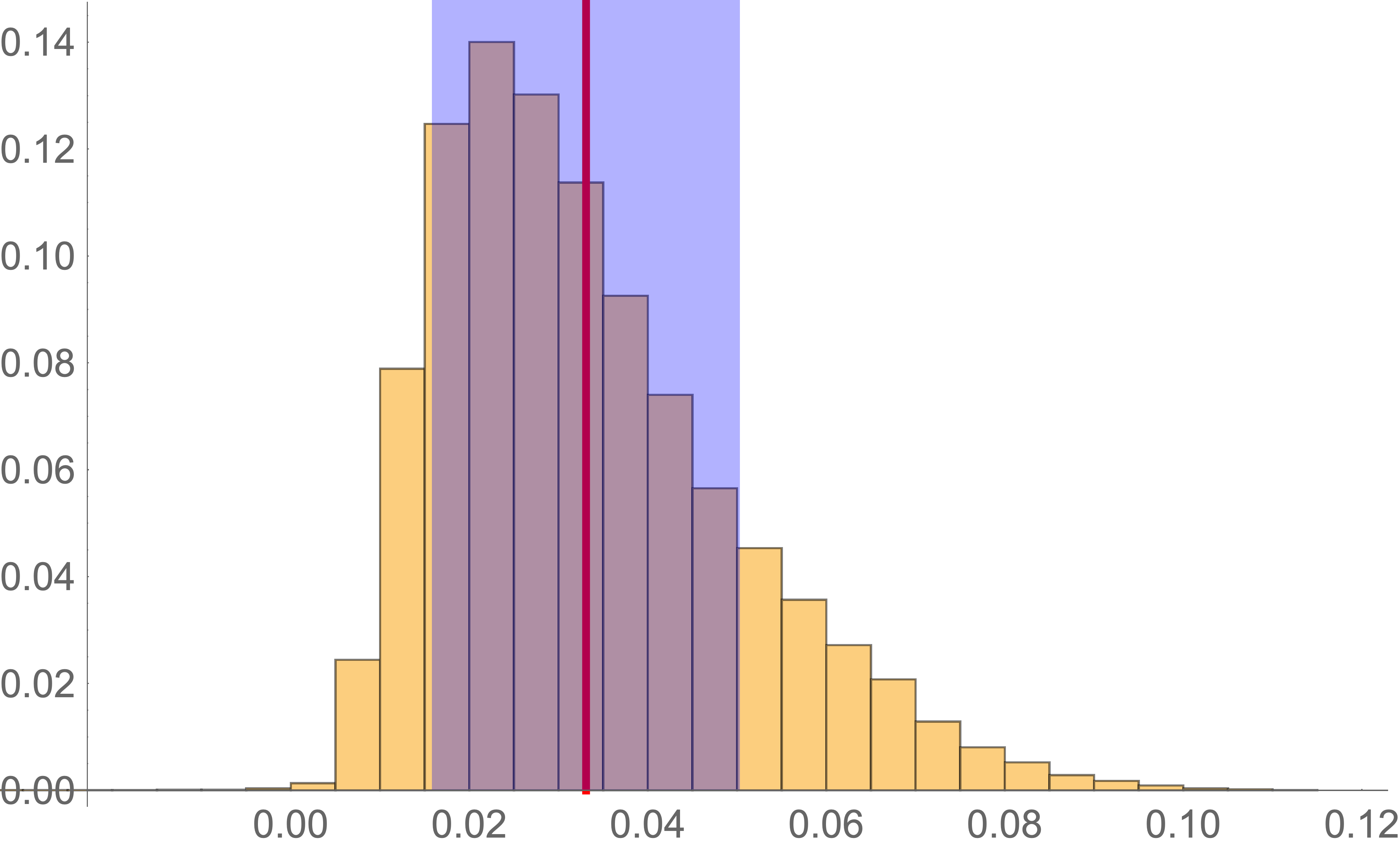}  
    \caption{$n=10^3$, $m=3.3\times10^{-2}$, $s=1.7\times10^{-2}$}
\end{subfigure}
\begin{subfigure}{.45\textwidth}
    \centering
    \includegraphics[width=\linewidth]{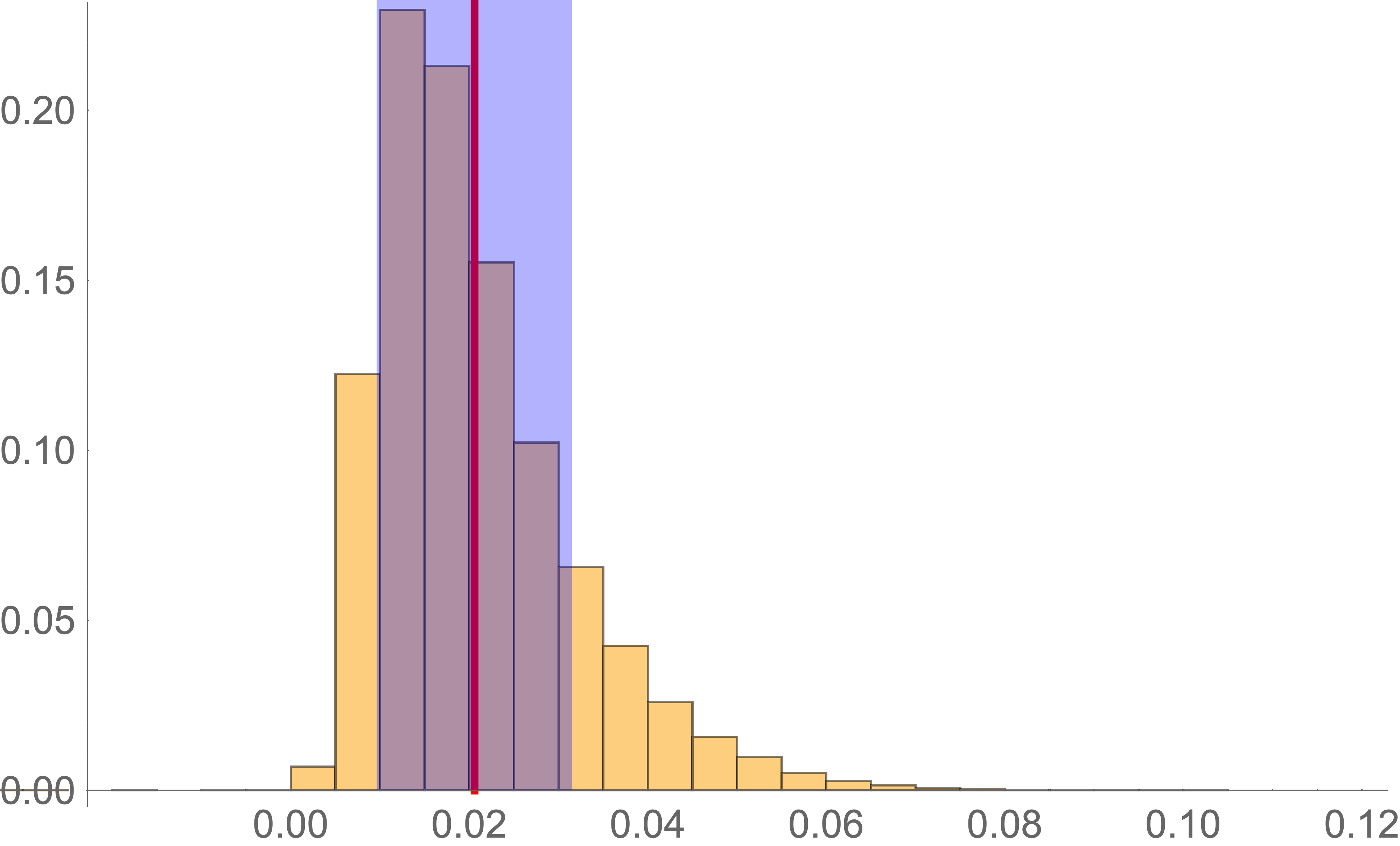}  
    \caption{$n=10^4$, $m=2.0\times10^{-2}$, $s=1.0\times10^{-2}$}
\end{subfigure}
\hfill
\begin{subfigure}{.45\textwidth}
    \centering
    \includegraphics[width=\linewidth]{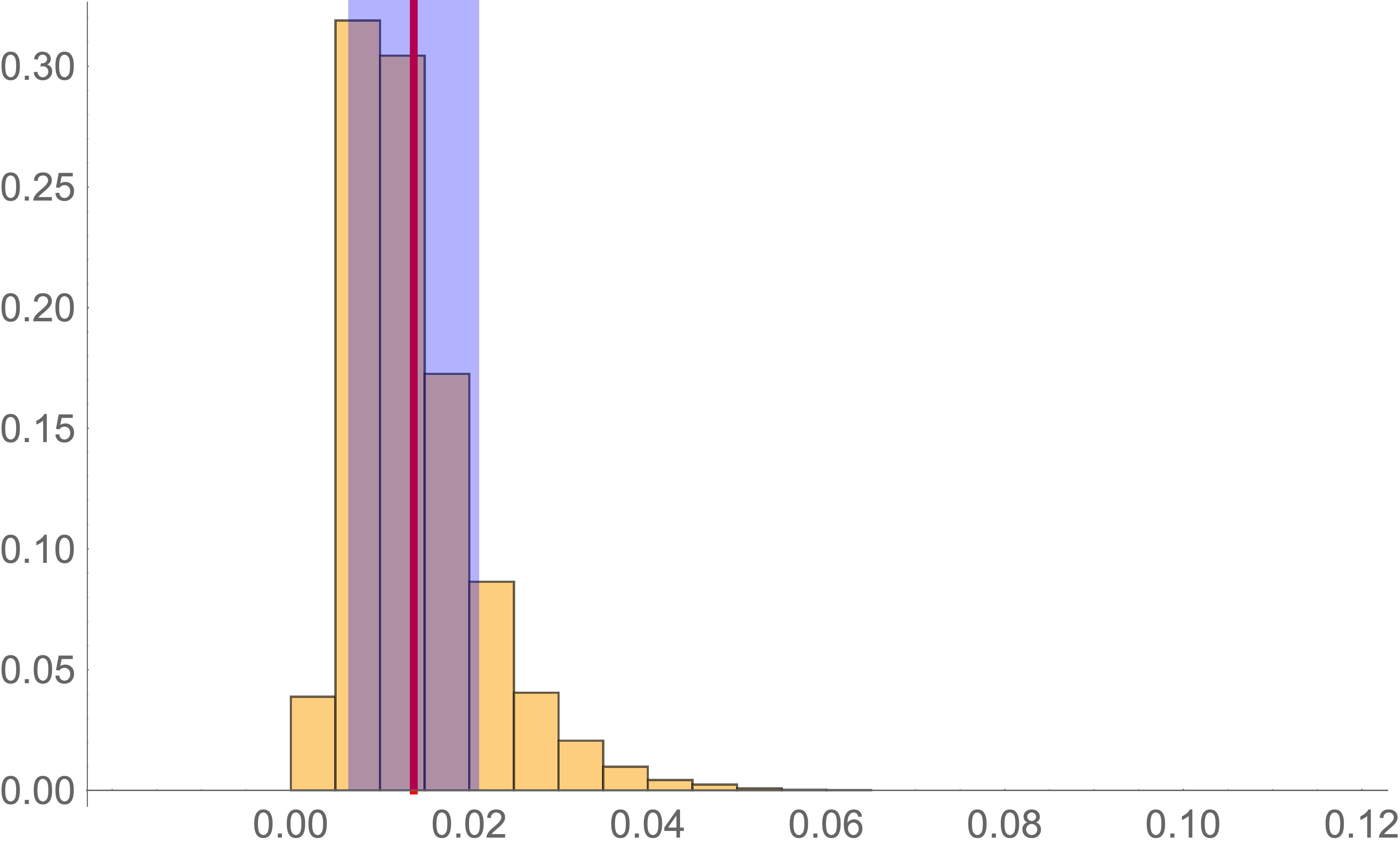}  
    \caption{$n=10^5$, $m=1.3\times10^{-2}$, $s=7.3\times10^{-3}$}
\end{subfigure}
\caption{Empirical distribution of $\lambda_{n,V}\norm{L_{n,V}}_\infty-\frac{\mu_1}{2}$ for $d=1$ and $V(0)\overset{\text{d}}{=}$ Bernoulli$(0.3)$ computed form $10^5$ samples. The empirical mean ($m$) and empirical standard deviation ($s$) are shown in red and blue respectively.}
\label{Fig2}
\end{figure}

Our second result is a partial proof of Conjecture \ref{Co}, and a complete proof when $d=1$.
\begin{Theorem}\label{Th2}\leavevmode
	\begin{itemize}
		\item[i)] $\displaystyle\varliminf_{n\to\infty}\lambda_{n,V}\norm{L_{n,V}}_\infty\geq\frac{\mu_d}{2d}$\quad$\mathbb{P}$-a.s.
		\item[ii)] If $d=1$ then
		$\displaystyle\lim_{n\to\infty}\lambda_{n,V}\norm{L_{n,V}}_\infty=\frac{\mu_1}{2}$\quad$\mathbb{P}$-a.s.
	\end{itemize}
\end{Theorem}
\begin{Remark}The preprint \cite{Chenn} has a proof of ii) in the continuous setting for the \textbf{(C1)} case. Both proofs follow the heuristic of the largest ball with zero or very low potential, but differ on how to obtain the lower bound of $\lambda_{n,V}$ and the upper bound of $L_{n,V}$.
\end{Remark}
We prove Theorem \ref{Th2} in Section 3 after deriving some general properties of landscape functions. Most notable among these properties is Proposition \ref{PrEVLF}, which states that $\lambda_{A,W}\norm{L_{A,W}}_\infty$ is bounded form above and bellow by two dimensional constants uniformly on $A$ and $W$. This is a consequence of an upper bound of the $\ell^\infty\to\ell^\infty$ norm of the semigroup generated by the Schrödinger operator, which we adapted from the book \cite{Sznitman} to the discrete setting. The statement i) of Theorem \ref{Th2} follows from domain monotonicity of the landscape function and the asymptotic of $\lambda_{n,V}$ given in Theorem \ref{Th1}, while ii) is based on the geometric resolvent identity and the restrictions of one dimensional geometry. In Figure \ref{Fig2} we illustrate ii) by showing the convergence of the empirical distribution of $\lambda_{n,V}\norm{L_{n,V}}_\infty-\frac{\mu_1}{2}$ towards $0$.

In the proofs that follow, $C(d)$ is a finite positive constant that may only depend on the dimension and can change form line to line. By $a_t\sim_t b_t$ we mean $\lim_{t\to\infty}\frac{a_t}{b_t}=1$.

\section{Principal Eigenvalue (Proof of Theorem \ref{Th1})}
\subsection{Upper Bound of $\lambda_{n,V}$}
We introduce the sequences
\begin{equation*}
	\epsilon_n\coloneqq\begin{cases}
		0,&\textbf{(C1)},\\
		(\ln n)^{-2/d},&\textbf{(C2)},
	\end{cases}
	\qquad y_n\coloneqq\left(\frac{d\ln n}{\omega_d\abs{\ln F(\epsilon_n)}}\right)^{1/d},
\end{equation*}
so we can write the goal of this subsection as
\begin{equation}\label{EqLSEV}
	\varlimsup_{n\to\infty}y_n^2\lambda_{n,V}\leq\mu_d\quad\mathbb{P}\text{-a.s.}
\end{equation}
As usual, getting a sharp upper bound on $\lambda_{n,V}$ is much easier than a sharp lower bound. It just requires choosing a good test function and applying the Min-Max Principle.

Let $Y_n$ be the radius of the largest open euclidean ball contained in $\Lambda_n$ in which $V$ is uniformly bounded by $\epsilon_n$, that is,
\begin{equation*}
	Y_n\coloneqq\max\left\{r\in\N\,\middle|\,\exists x\in\Lambda_n\,\text{ such that }\,B(x,r)\cap\Z^d\subseteq\Lambda_n\cap V^{-1}\big([0,\epsilon_n]\big)\right\},
\end{equation*}
where $B(x,r)=\{x'\in\R^d\,|\,\abs{x-x'}<r\}\subseteq\R^d$. Also, let $x_n\in\Lambda_n$ be the center of a ball at which the maximum is attained (it may not be unique). The asymptotic growth of $Y_n$ is given in the next proposition, whose proof we delay a short moment.
\begin{Proposition}\label{PrY}
	$\displaystyle Y_n\sim_n y_n$\quad$\mathbb{P}$-a.s.
\end{Proposition}
Let $\phi\in\ell^2(B(x_n,Y_n)\cap\Z^d)$ be the normalized eigenvector of $-\Delta_{B(x_n,Y_n)}$ associated to $\lambda_{B(x_n,Y_n)\cap\Z^d,0}$ and extend it by $0$ to $\Lambda_n$. Then, by the Min-Max Principle, we have
\begin{align*}
    \lambda_{n,V}&\leq\hp{\phi}{\left(-\Delta_{n}+V\right)\phi}_{\ell^2(\Lambda_n)}=\hp{\phi}{\left(-\Delta_{B(x_n,Y_n)\cap\Z^d}+V\right)\phi}_{\ell^2(B(x_n,Y_n)\cap\Z^d)}\\&\leq \lambda_{B(x_n,Y_n)\cap\Z^d,0}+\epsilon_n
\end{align*}
and therefore
\begin{equation*}
	\varlimsup_{n\to\infty}y_n^2\lambda_{n,V}\leq \lim_{n\to\infty}y_n^2\left(\lambda_{B(x_n,Y_n)\cap\Z^d,0}+\epsilon_n\right)=\lim_{n\to\infty}\frac{y_n^2}{Y_n^2}Y_n^2\lambda_{B(x_n,Y_n)\cap\Z^d,0}=\mu_d\quad\mathbb{P}\text{-a.s.},
\end{equation*}
where we have used Proposition \ref{PrY}, $\displaystyle\lim_{r\to\infty}r^2\lambda_{B(0,r)\cap\Z^d,0}=\mu_d$ and translation invariance. This last limit is a consequence of the discrete Laplacian converging to the continuous one, or random walk converging to Brownian motion. A proof following the latter approach can be found in \cite[Proposition 8.4.2]{Limic}, where an extra factor $d$ appears as a result of the probabilistic normalization of the Laplacian.
\begin{proof}[Proof of Proposition \upshape{\ref{PrY}}]
	If $Y_n< y_n(1-\delta)^{1/d}$ for some $0<\delta<1$, then the inscribed ball of each of the $\left(\frac{2n}{2 y_n(1-\delta)^{1/d}}\right)^d(1+o(1))$ disjoint cubes, of side length $\ceil{2 y_n(1-\delta)^{1/d}}$, that make up $\Lambda_n$ contains a point $x$ with $V(x)>\epsilon_n$. Approximating the number of points in such balls by $\#(B(0,r)\cap\Z^d)\sim_r\vol(B(0,r))=\omega_d r^d$, we obtain for large $n$
	\begin{align*}
		\PPrb{Y_n< y_n(1-\delta)^{1/d}}&\leq\left(1-F(\epsilon_n)^{\omega_d y_n^d(1-\delta)(1+o(1))}\right)^{\frac{n^d}{y_n^d(1-\delta)}(1+o(1))}\\
		&=\left(1-\frac{1}{n^{d(1-\delta)(1+o(1))}}\right)^{\frac{n^d\omega_d\abs{\ln F(\epsilon_n)}}{d(\ln n)(1-\delta)}(1+o(1))}\\
		&\leq \exp\left(-\frac{n^\delta\omega_d\abs{\ln F(\epsilon_n)}}{2 d(\ln n)(1-\delta)}\right),
	\end{align*}
	which is summable. Therefore, the Borel–Cantelli Lemma and sending $\delta\to 0$ give
	\begin{equation*}
		1\leq\varliminf_{n\to\infty}y_n^{-1} Y_n\quad\mathbb{P}\text{-a.s.}
	\end{equation*}
	We show the $\limsup$ bound first on an exponential sub-sequence and then we extend it to the whole sequence. The extending argument requires a monotone sequence of random variables, which $Y_n$ may fail to be if \textbf{(C2)} holds. For this reason we introduce
	\begin{equation*}
		Y_{n,n'}\coloneqq\max\left\{r\in\N\,\middle|\,\exists x\in\Lambda_n\,\text{ such that }\,B(x,r)\cap\Z^d\subseteq\Lambda_n\cap V^{-1}\big([0,\epsilon_{n'}]\big)\right\},
	\end{equation*}
	which is increasing on $n$, decreasing on $n'$ and satisfies $Y_{n,n}=Y_n$. Since for $\delta>0$ and large $m$ we have
	\begin{align*}
		\mathbb{P}\left[Y_{\floor{e^{m+1}},\floor{e^{m}}}> y_{\floor{e^{m}}}(1+\delta)^{1/d}\right]&\leq\sum_{x\in\Lambda_{\floor{e^{m+1}}}}\mathbb{P}\left[B(x,y_{\floor{e^{m}}}(1+\delta)^{1/d})\cap\Z^d\subseteq V^{-1}[0,\epsilon_{\floor{e^m}}]\right]\\
		&=\#\Lambda_{\floor{e^{m+1}}} F(\epsilon_{\floor{e^m}})^{\omega_d y_{\floor{e^{m}}}^d(1+\delta)(1+o(1))}\\
		&= \frac{\#\Lambda_{\floor{e^{m+1}}}}{\floor{e^{m}}^{d(1+\delta)(1+o(1))}}\\
		&\leq C(d) e^{-m d \delta/2},
	\end{align*}
	the Borel–Cantelli Lemma and the limit $\delta\to 0$ give 
	\begin{equation*}
		\varlimsup_{m\to\infty}y_{\floor{e^{m}}}^{-1}Y_{\floor{e^{m+1}},\floor{e^{m}}}\leq1\quad\mathbb{P}\text{-a.s.}
	\end{equation*}
	For $n\in\N$ define $m(n)\in\N$ by $\floor{e^{m(n)}}\leq n<\floor{e^{m(n)+1}}$. Since $y_n\sim_n y_{\floor{e^{m(n)}}}$ and $Y_n\leq Y_{\floor{e^{m(n)+1}},\floor{e^{m(n)}}}$ we conclude
	\begin{equation*}
		\varlimsup_{n\to\infty}y_n^{-1}Y_n\leq \varlimsup_{n\to\infty}y_n^{-1} Y_{\floor{e^{m(n)+1}},\floor{e^{m(n)}}}=\varlimsup_{n\to\infty}y_{\floor{e^{m(n)}}}^{-1}Y_{\floor{e^{m(n)+1}},\floor{e^{m(n)}}}\leq1\quad\mathbb{P}\text{-a.s.}\qedhere
	\end{equation*}
\end{proof}

\subsection{Lower Bound of $\lambda_{n,V}$}
In this subsection we show that $\mathbb{P}$-a.s. we have
\begin{equation}\label{EqLIEV}
    1\leq\varliminf_{n\to\infty}\frac{\lambda_{n,V}}{\mu_d\left(\frac{\omega_d\abs{\ln F(0)}}{d\ln n}\right)^{2/d}}\qquad\text{and}\qquad  1\leq\varliminf_{n\to\infty}\frac{\lambda_{n,V}}{\mu_d\left(\frac{2\eta \omega_d\ln \ln n}{d^2\ln n}\right)^{2/d}}
\end{equation}
for \textbf{(C1)} and \textbf{(C2)} respectively. The main input for this is a  Lifshitz tail result on the integrated density of states from \cite{Biskup}. We recall the integrated density of states of the Anderson model is a deterministic distribution function given by the $\mathbb{P}$-a.s. limit
\begin{equation*}
    I(t)\coloneqq\lim_{n\to\infty}\frac{1}{\#\Lambda_n}\#\{\lambda\in\sigma(-\Delta_n+V)\,|\,\lambda\leq t\},\qquad t\in\R,
\end{equation*}
where the eigenvalues are counted with multiplicities. The central hypothesis of \cite{Biskup} is a scaling assumption of the cumulant-generating function $H(t)\coloneqq\ln\EE{e^{-t V(0)}}$ of $V(0)$, which we prove in the following proposition. To state it, we first need to define
\begin{equation*}
(1,\infty)\ni t\longmapsto\alpha(t)\coloneqq
\begin{cases}t^{1/(d+2)},& \textbf{(C1)},\\
\left(\frac{t}{\ln t}\right)^{1/(d+2)},&\textbf{(C2)},
\end{cases}\qquad \widetilde{H}\coloneqq
\begin{cases}\abs{\ln F(0)},& \textbf{(C1)},\\
\frac{2\eta}{d+2},& \textbf{(C2)}.
\end{cases}
\end{equation*}
\begin{Proposition}
For any compact $K\subseteq(0,\infty)$ we have
\begin{equation*}
    \lim_{t\to\infty}\frac{\alpha^{d+2}(t)}{t}H\left(\frac{t}{\alpha^d(t)}y\right)=-\widetilde{H}
\end{equation*}
uniformly on $y\in K$.
\end{Proposition}
\begin{proof}
First assume \textbf{(C1)}. In this case $\frac{\alpha^{d+2}(t)}{t}=1$ and $\frac{t}{\alpha^d(t)}=t^{2/(d+2)}$. Since for $t>0$ we have
\begin{align*}
    \ln F(0)\leq H(t)&=\ln\left(\EE{e^{-t V(0)}\1_{V(0)\leq \frac{1}{\sqrt{t}}}}+\EE{e^{-t V(0)}\1_{V(0)> \frac{1}{\sqrt{t}}}}\right)\\
    &\leq \ln\left(F\left(1/\sqrt{t}\right)+e^{-\sqrt{t}}\right),
\end{align*}
we conclude that
\begin{align*}
    \sup_{y\in K}\abs{\frac{\alpha^{d+2}(t)}{t}H\left(\frac{t}{\alpha^d(t)}y\right)-\ln F(0)}&\leq \sup_{y\in K}\ln\left(F\left(\frac{1}{t^{1/d+2}\sqrt{y}}\right)+e^{-t^{1/d+2}\sqrt{y}}\right)-\ln F(0)\\
    &=\ln\left(F\left(\frac{1}{t^{1/d+2}\sqrt{\min K}}\right)+e^{-t^{1/d+2}\sqrt{\min K}}\right)-\ln F(0)\\
    &\quad\xrightarrow[t\to \infty]{} 0.
\end{align*}

Now assume \textbf{(C2)}. In this case $\frac{\alpha^{d+2}(t)}{t}=\frac{1}{\ln t}$ and $\frac{t}{\alpha^d(t)}=t^{2/(d+2)}(\ln t)^{d/(d+2)}$. We introduce a parameter $0<\delta<1$ and observe
\begin{equation*}
    H(t)=\ln\left(\EE{e^{-t V(0)}\1_{V(0)\leq t^{-\delta}}}+\EE{e^{-t V(0)}\1_{V(0)>t^{-\delta}}}\right)\leq \ln\left(F\left(t^{-\delta}\right)+e^{-t^{1-\delta}}\right),\qquad t>0,
\end{equation*}
which implies
\begin{align*}
    &\varlimsup_{t\to\infty}\sup_{y\in K}\frac{\alpha^{d+2}(t)}{t}H\left(\frac{t}{\alpha^d(t)}y\right)\\
    &\qquad\leq \varlimsup_{t\to\infty}\sup_{y\in K}\frac{1}{\ln t}\ln\left(F\left(\left[\frac{t}{\alpha^d(t)}y\right]^{-\delta}\right)+\exp\left(-\left[\frac{t}{\alpha^d(t)}y\right]^{1-\delta}\right)\right)\\
    &\qquad= \varlimsup_{t\to\infty}\frac{1}{\ln t}\ln\left(F\left(\left[\frac{t}{\alpha^d(t)}\min K\right]^{-\delta}\right)+\exp\left(-\left[\frac{t}{\alpha^d(t)}\min K\right]^{1-\delta}\right)\right)\\
    &\qquad=-\frac{2\delta\eta}{d+2}\xrightarrow[\delta\to1]{}-\frac{2\eta}{d+2}.
\end{align*}
For the $\displaystyle\varliminf_{t\to\infty}\inf_{y\in K}$ we use
\begin{equation*}
    H(t)=\ln\left(\EE{e^{-t V(0)}\1_{V(0)\leq t^{-1}}}+\EE{e^{-t V(0)}\1_{V(0)>t^{-1}}}\right)\geq\ln\left(e^{-1}F(t^{-1})\right),\qquad t>0
\end{equation*}
to obtain
\begin{align*}
    \varliminf_{t\to\infty}\inf_{y\in K}\frac{\alpha^{d+2}(t)}{t}H\left(\frac{t}{\alpha^d(t)}y\right)&\geq \varliminf_{t\to\infty}\inf_{y\in K}\frac{1}{\ln t}\ln\left(e^{-1}F\left(\left[\frac{t}{\alpha^d(t)}y\right]^{-1}\right)\right)\\
    &\geq \varliminf_{t\to\infty}\frac{1}{\ln t}\ln\left(e^{-1}F\left(\left[\frac{t}{\alpha^d(t)}\max K\right]^{-1}\right)\right)=-\frac{2\eta}{d+2}.\qedhere
\end{align*}
\end{proof}
Having checked the scaling assumption on $H$, we now have the Lifshitz tail result:
\begin{Theorem}[Theorem 1.3 of \cite{Biskup}]\label{ThLT}
Define $\displaystyle\chi\coloneqq\inf_{g\in H^1(\R^d),\,\norm{g}_2=1}\left(\norm{\nabla g}_2^2+\widetilde{H}\vol(\supp g)\right)$, then
\begin{equation*}
    \lim_{t\downarrow0}\frac{\ln I(t)}{t\alpha^{-1}\left(t^{-1/2}\right)}=-2\,d^{d/2}\left(\frac{\chi}{d+2}\right)^{(d+2)/2}.
\end{equation*}

\end{Theorem}
\begin{Remark}
The function $t\mapsto\alpha(t)$ is eventually increasing so $\alpha^{-1}(t)$ is well defined for large $t$. The original statement from \cite{Biskup} is far more general; our conditions on $V$ make $H$ fall into, what is there called, the $(\gamma=0)$-class.
\end{Remark}
The constant $\chi$ can be explicitly computed by means of the Faber-Krahn inequality:
\begin{Proposition}\label{PrCHI}
$\displaystyle\chi=(d+2)\left(\frac{\widetilde{H}\omega_d}{2}\right)^{2/(d+2)}\left(\frac{\mu_d}{d}\right)^{d/(d+2)}$.
\end{Proposition}
\begin{proof}
Starting from $\chi=\inf_{g\in H^1(\R^d),\,\norm{g}_2=1}\left(\norm{\nabla g}_2^2+D\vol(\supp g)\right)$ we see that we only need to consider the finite volume case. Hence
\begin{equation*}
    \chi=\inf_{\substack{A\subseteq\R^d,\\\vol(A)<\infty}}\inf_{\substack{g\in H^1(\R^d),\,\norm{g}_2=1,\\\supp g=A}}\left(\norm{\nabla g}_2^2+\widetilde{H}\vol(A)\right)=\inf_{\substack{A\subseteq\R^d,\\\vol(A)<\infty}}\left(\mu(A)+\widetilde{H}\vol(A)\right),
\end{equation*}
where $\mu(A)$ is the principal eigenvalue of the continuous Laplacian ($-\sum_{i=1}^d\partial^2/\partial x_i^2$) defined on $A$ with Dirichlet boundary conditions. The Faber-Krahn inequality states that over all domains of a given volume the one with the lowest principal eigenvalue is the ball, therefore, using $\mu(B(0,r))=\mu_d/r^2$ and $\vol(B(0,r))=\omega_d r^d$ we obtain
\begin{equation*}
  \chi=\inf_{0<r<\infty}\left(\frac{\mu_d}{r^2}+\widetilde{H}\omega_d r^d\right).  
\end{equation*}
Evaluating at the only critical point $\displaystyle r=\left(\frac{2\mu_d}{\widetilde{H}\omega_d d}\right)^{1/(d+2)}$ finishes the proof.
\end{proof}
We now exploit the connection between $I$ and the distribution of $\lambda_{n,V}$. This is a classic argument that can be found, for instance, in \cite[Equation 4.46]{Aizenman}. We present here a slightly modified version. Let $n\in\N$ and define a new potential
\begin{equation*}
    V'(x)\coloneqq\begin{cases}
    \infty,& x\in(2n+2)\Z^d,\\
    V(x),&\text{else}.
    \end{cases}
\end{equation*}
Clearly $V\leq V'$ so for any $k\in\N$ and $t\in\R$ we have
\begin{equation*}
    \frac{\#\{\lambda\in\sigma(-\Delta_{(2n+2)k}+V)\,|\,\lambda\leq t\}}{\#\Lambda_{(2n+2)k}}\geq \frac{\#\{\lambda\in\sigma(-\Delta_{(2n+2)k}+V')\,|\,\lambda\leq t\}}{\#\Lambda_{(2n+2)k}}
\end{equation*}
where we use implicitly the convention of Dirichlet boundary conditions wherever $V'$ is infinite. Taking $k\to\infty$ and noting that the infinities of $V'$ decompose $-\Delta_{(2n+2)k}+V'$ in to a direct sum of $(2k)^d$ independent terms equal in distribution to $-\Delta_{n}+V$ we obtain
\begin{align*}
    I(t)&\geq \left(\lim_{k\to\infty}\frac{(2k)^d}{\#\Lambda_{(2n+2)k}}\right)\EE{\#\{\lambda\in\sigma(-\Delta_{n}+V)\,|\,\lambda\leq t\}}\\
    &\geq\left(\frac{1}{2n+2}\right)^d\PPrb{\lambda_{n,V}\leq t}.
\end{align*}
From the previous inequality, Theorem \ref{ThLT} and Proposition \ref{PrCHI} we have
\begin{equation}\label{EqLBEV}
\PPrb{\lambda_{n,V}\leq t}\leq C(d) n^{d} I(t)\leq C(d)\,n^{d}\exp\left[-f(1/t)(1+o(1))\right]\quad\text{as } t\downarrow0,
\end{equation}
where we have introduced $\displaystyle f(t)\coloneqq\frac{\widetilde{H}\omega_d\mu_d^{d/2}\alpha^{-1}(t^{1/2})}{t}$. To finish the proof we need the asymptotic of $f^{-1}(t)$ as $t\to\infty$:
\begin{Proposition}\label{PrF}\leavevmode
\begin{itemize}
    \item For \textbf{(C1)}, $\displaystyle f^{-1}(t)=\frac{1}{\mu_d}\left(\frac{t}{\omega_d\abs{\ln F(0)}}\right)^{2/d}$.
    \item For \textbf{(C2)}, $\displaystyle f^{-1}(t)\sim_t\frac{1}{\mu_d}\left(\frac{d\,t}{2\eta\omega_d\ln t}\right)^{2/d}$.
\end{itemize}
\end{Proposition}
\begin{proof}
For \textbf{(C1)} there is nothing to prove since $f(t)=\omega_d\abs{\ln F(0)}\mu_d^{d/2}t^{d/2}$. For \textbf{(C2)} we have
\begin{equation*}
    f(t)=\frac{k\alpha^{-1}(t^{1/2})}{t}=k t^{d/2}\ln \alpha^{-1}(t^{1/2}),
\end{equation*}
with all the constants collected in $k=\frac{2\eta \omega_d\mu_d^{d/2}}{d+2}$. Since $\alpha$ is eventually increasing and has infinite limit, the same is true for $f$, in particular $f^{-1}(t)$ exists for large $t$. By solving for the $\alpha^{-1}$ term in the first equality above, applying $\alpha$ and simplifying some exponents we arrive at $t=\left(\frac{f(t)}{k\ln\left[t f(t)/k\right]}\right)^{2/d}$. Replacing $t$ by $f^{-1}(t)$ we obtain
\begin{equation*}
    f^{-1}(t)=\left(\frac{t}{k\ln\left[t f^{-1}(t)/k\right]}\right)^{2/d},
\end{equation*}
and then multiplying by $t$ and taking the logarithm leads to
\begin{equation*}
    \ln [t f^{-1}(t)]=\frac{d+2}{d}\ln t-\frac{2}{d}\ln\left[k\ln\left[t f^{-1}(t)/k\right]\right],
\end{equation*}
which implies 
\begin{equation*}
     f^{-1}(t)\sim_t\left(\frac{d\,t}{(d+2)k\ln t}\right)^{2/d}=\frac{1}{\mu_d}\left(\frac{d\,t}{2\eta\omega_d\ln t}\right)^{2/d}.\qedhere
\end{equation*}
\end{proof}
Going back to \eqref{EqLBEV} with $n=\floor{e^m}$ and $t=1/f^{-1}((1+\delta)d m)$ for some $m\in\N$ and $\delta>0$, we see that
\begin{align*}
    \PPrb{\lambda_{\floor{e^m},V}f^{-1}((1+\delta)d m)\leq 1}&\leq C(d) \left(\floor{e^m}\right)^{d}\exp\left[-(1+\delta)d m(1+o(1))\right]\\&\leq C(d)\,e^{-m d\delta/2},
\end{align*}
which is summable over $m\in\N$. Therefore, by the Borel–Cantelli Lemma we have
\begin{equation*}
    1\leq \varliminf_{m\to\infty} \lambda_{\floor{e^m},V}f^{-1}((1+\delta)d m)=(1+\delta)^{2/d} \varliminf_{m\to\infty} \lambda_{\floor{e^m},V}f^{-1}(d m)\quad\mathbb{P}\text{-a.s.}
\end{equation*}
As in the proof of Proposition \ref{PrY}, we define $m(n)\in\N$ by $\floor{e^{m(n)}}\leq n<\floor{e^{m(n)+1}}$, so that $\ln n\sim_n (m(n)+1)$. Since $n\mapsto\lambda_{n,V}$ is monotone decreasing we have
\begin{align*}
\varliminf_{n\to\infty} \lambda_{n,V}f^{-1}(d \ln n)&\geq \varliminf_{n\to\infty} \lambda_{\floor{e^{m(n)+1}},V}f^{-1}(d \ln n)\\&=\varliminf_{n\to\infty} \lambda_{\floor{e^{m(n)+1}},V}f^{-1}(d(m(n)+1))\geq (1+\delta)^{-2/d}\quad\mathbb{P}\text{-a.s.}
\end{align*}
By sending $\delta\to 0$ and replacing the $f^{-1}$ term by its asymptotic given in Proposition \ref{PrF} we obtain the desired result of this subsection.

\section{Landscape Function}
We start this section by deriving some general properties of landscape functions. 

For a finite $A\subseteq\Z^d$ and $W:A\rightarrow[0,\infty)$ we introduce the Green function (with $0$ as spectral parameter) 
\begin{equation*}
    G_{A,W}(x,y)\coloneqq\begin{cases}
    \hp{\delta_x}{(-\Delta_A+W)^{-1}\delta_y}_{\ell^2(A)},&(x,y)\in A\times A,\\0,&(x,y)\in (\Z^d\times\Z^d)\setminus(A\times A).
    \end{cases}
\end{equation*}
This function is known to be symmetric, non-negative, decreasing on the potential $W$; and to satisfy the geometric resolvent identity (see \cite[Section 5.3]{Kirsch}): if $A'\subseteq A$ then
\begin{equation*}
    G_{A,W}(x,y)=G_{A',W}(x,y)+\sum_{(i,j)\in\partial A'}G_{A',W}(x,i)G_{A,W}(j,y),
\end{equation*}
where $\partial A'\coloneqq\{(i,j)\in\Z^d\times\Z^d\,|\,i\in A',\,j\notin A',\,\abs{i-j}=1\}$ is the boundary of $A'$. By extending the definition of $L_{A,W}$ to $L_{A,W}(x)\coloneqq\sum_{y\in \Z^d}G_{A,W}(x,y)$ for all $x\in\Z^d$, the previously stated properties of $G_{A,W}$ translate into non-negativity, potential monotonicity and domain monotonicity of landscape functions:
\begin{itemize}
    \item $L_{A,W}\geq0$ and $L_{A,W}(x)=0$ if $x\in\Z^d\setminus A$.
    \item If $0\leq W'\leq W$ then $L_{A,W}\leq L_{A,W'}$.
    \item If $A'\subseteq A$ then 
    \begin{equation}\label{EqGRI}
        L_{A,W}(x)=L_{A',W}(x)+\sum_{(i,j)\in\partial A'}G_{A',W}(x,i)L_{A,W}(j)\geq L_{A',W}(x).
    \end{equation}
\end{itemize}

Our last general property is that $\lambda_{A,W}\norm{L_{A,W}}_\infty$ is bounded from above and bellow by two positive constants uniformly on $A$ and $W$. This is based on the following upper bound of the $\ell^\infty\to\ell^\infty$ norm of the semigroup, which can be found, for the continuous setting, in \cite[Chapter 3, Theorem 1.2]{Sznitman}. We could not find a proof in the literature for the discrete case, so we provide one in Appendix A.
\begin{Theorem}\label{ThSG}
For any finite $A\subseteq\Z^d$ and $W:A\rightarrow[0,\infty)$ we have
\begin{equation*}
    \norm{\exp\left(-t[-\Delta_{A}+W]\right)\1_A}_\infty\leq C(d)\left (1+\left[\lambda_{A,W}t\right]^{d/2}\right)\exp\left(-\lambda_{A,W}t\right),\qquad t\geq0.
\end{equation*}
\end{Theorem}
As an an immediate consequence we obtain:
\begin{Proposition}\label{PrEVLF}
For any finite $A\subseteq\Z^d$ and $W:A\rightarrow[0,\infty)$ we have
\begin{equation*}
    1\leq\lambda_{A,W}\norm{L_{A,W}}_\infty\leq C(d).
\end{equation*}
\end{Proposition}
\begin{Remark}
The lower bound is sharp. It is attained when $A$ is a single point of $\Z^d$. 
\end{Remark}
\begin{proof}
For the upper bound we use Theorem \ref{ThSG} and the substitution $u=\lambda_{A,W}t$:
\begin{align*}
    \norm{L_{A,W}}_\infty&= \norm{\int_0^\infty\exp\left(-t[-\Delta_{A}+W]\right)\1_{A}\dd{t}}_\infty\\
    &\leq C(d)\int_0^\infty\left(1+\left[\lambda_{A,W}t\right]^{d/2}\right)\exp\left(-\lambda_{A,W}t\right)\dd{t}\\
    &=\frac{C(d)}{\lambda_{A,W}}\int_0^\infty\left(1+u^{d/2}\right)e^{-u}\dd{u}=\frac{C(d)}{\lambda_{A,W}}.
\end{align*}
For the lower bound we just need to notice that the positivity of $G_{A,W}$ implies 
\begin{equation*}
\sup_{\phi\in\ell^2(A)\setminus\{0\}}\frac{\norm{(-\Delta_A+W)^{-1}\phi}_\infty}{\norm{\phi}_\infty}=\sup_{\phi\in\ell^2(A)\setminus\{0\}}\frac{\norm{\sum_{y\in A}G_{A,W}(\cdot,y)\phi(y)}_\infty}{\norm{\phi}_\infty}=\norm{L_{A,W}}_\infty.
\end{equation*}
By plugging in the eigenvector associated to $\lambda_{A,W}$ we obtain $\norm{L_{A,W}}_\infty\geq\frac{1}{\lambda_{A,W}}$.
\end{proof}

\subsection{Proof of Theorem \ref{Th2} i)}
We start with the asymptotic of the sup-norm of the landscape function on balls with $0$ potential.
\begin{Proposition}\label{PrLFB}
$\displaystyle\norm{L_{B(0,r)\cap \Z^d,0}}_\infty\sim_r\frac{r^2}{2d}$.
\end{Proposition}
\begin{proof}
Let $r>0$ and consider the function $\phi_r(x)\coloneqq\frac{r^2-\abs{x}^2}{2d}$ defined on $\Z^d$. Clearly $-\Delta\phi_r(x)=1$ for all $x\in\Z^d$ and therefore $L_{B(0,r)\cap \Z^d,0}-\phi_r$ is harmonic in $B(0,r)\cap \Z^d$. By the Maximum Principle we have
\begin{align*}
    \abs{\norm{L_{B(0,r)\cap \Z^d,0}}_\infty-\frac{r^2}{2d}}&=\abs{\sup_{x\in B(0,r)\cap \Z^d}L_{B(0,r)\cap \Z^d,0}(x)-\sup_{x\in B(0,r)\cap \Z^d}\phi_r(x)}\\
    &\leq\sup_{x\in B(0,r)\cap \Z^d}\abs{L_{B(0,r)\cap \Z^d,0}(x)-\phi_r(x)}\\&=\sup_{x\in \partial^+\left[B(0,r)\cap \Z^d\right]}\abs{L_{B(0,r)\cap \Z^d,0}(x)-\phi_r(x)}\\&=\sup_{x\in \partial^+\left[B(0,r)\cap \Z^d\right]}\abs{\phi_r(x)}\leq C(d)\,r(1+o(1))
\end{align*}
where $\partial^+ A\coloneqq\left\{x\in \Z^d\setminus A\,\middle|\,\exists\,y\in A\text{ such that }\abs{x-y}=1\right\}$ is the outer boundary of $A\subseteq\Z^d$. Dividing by $\frac{r^2}{2d}$ and taking the limit $r\to\infty$ give the proposition. 
\end{proof}

Recall the definitions of $\epsilon_n$, $Y_n$, $x_n$ and $y_n$ from Subsection 2.1 and notice that Theorem \ref{Th1} can be restated as $\lambda_{n,V}\sim_n\frac{\mu_d}{y_n^2}$\quad$\mathbb{P}$-a.s. From domain monotonicity of landscape functions we have
\begin{equation*}
    L_{n,V}\geq L_{B(x_n,Y_n)\cap\Z^d,V}.
\end{equation*}
For \textbf{(C1)}, $V$ is identically $0$ in $B(x_n,Y_n)\cap\Z^d$ so Theorem \ref{Th1}, Proposition \ref{PrLFB} and translation invariance give
\begin{equation*}
    \varliminf_{n\to\infty}\lambda_{n,V}\norm{L_{n,V}}_\infty\geq  \lim_{n\to\infty}\lambda_{n,V}\norm{L_{B(x_n,Y_n)\cap \Z^d,0}}_\infty=\lim_{n\to\infty}\frac{\mu_d}{y_n^2}\frac{Y_n^2}{2d}=\frac{\mu_d}{2d}\quad\mathbb{P}\text{-a.s.}
\end{equation*}
For \textbf{(C2)}, we use the second resolvent identity, domain monotonicity of the eigenvalue, and Propositions \ref{PrEVLF}, \ref{PrLFB}, \ref{PrY} to obtain
\begin{align*}
    \lambda_{n,V}\norm{L_{B(x_n,Y_n)\cap \Z^d,0}-L_{B(x_n,Y_n)\cap \Z^d,V}}_\infty&= \lambda_{n,V}\norm{(-\Delta_{B(x_n,Y_n)\cap \Z^d,0})^{-1}V L_{B(x_n,Y_n)\cap \Z^d,V}}_\infty\\
    &\leq C(d)\epsilon_n \norm{L_{B(x_n,Y_n)\cap \Z^d,0}}_\infty\\
    &\leq C(d)\epsilon_n Y_n^2\xrightarrow[n\to\infty]{\mathbb{P}\text{-a.s.}}0,
\end{align*}
which implies
\begin{align*}
    \varliminf_{n\to\infty}\lambda_{n,V}\norm{L_{n,V}}_\infty&\geq  \lim_{n\to\infty}\lambda_{n,V}\norm{L_{B(x_n,Y_n)\cap \Z^d,V}}_\infty\\&=\lim_{n\to\infty}\lambda_{n,V}\norm{L_{B(x_n,Y_n)\cap \Z^d,0}}_\infty=\frac{\mu_d}{2d}\quad\mathbb{P}\text{-a.s.}
\end{align*}
This concludes the proof of Theorem \ref{Th2} i).

\subsection{Proof of Theorem \ref{Th2} ii)}
We assume from this point on that $d=1$. We set $\II{a}{b}\coloneqq[a,b]\cap
\Z$ for any two $a,b\in\Z$. This proof is based on the following deterministic bound of the Green function in terms of the values of the potential.
\begin{Proposition}\label{PrGF}
Let $n\in\N$ and $W:\II{1}{n}\rightarrow[0,\infty)$. For any $y\in\II{1}{n}$ we have
\begin{align*}
    G_{\II{1}{n},W}(1,y)&\leq\left(\sum_{j=0}^{y-1} (y-j) W(y-j)\right)^{-1},\\ G_{\II{1}{n},W}(y,n)&\leq\left(\sum_{j=0}^{n-y} (n-y+1-j) W(y+j)\right)^{-1}.
\end{align*}
\end{Proposition}
\begin{proof}
We only prove the first inequality; the second one follows from reflecting $W$ across the midpoint of $\II{1}{n}$ and the symmetry of the Green function.

Fix some $y\in\II{1}{n}$. By potential monotonicity we have $G_{\II{1}{n},W}\leq G_{\II{1}{n},W\1_{\II{1}{y}}}$. The Cramer's rule lets us write
\begin{equation*}
G_{\II{1}{n},W\1_{\II{1}{y}}}(1,y)=\frac{\det\left([-\Delta_{\II{1}{n}}+W\1_{\II{1}{y}}]_{1\to\delta_y}\right)}{\det(-\Delta_{\II{1}{n}}+W\1_{\II{1}{y}})},
\end{equation*}
where $[-\Delta_{\II{1}{n}}+W\1_{\II{1}{y}}]_{1\to\delta_y}$ is the matrix obtained by replacing the first column (in the canonical $\delta_j$ basis) of $-\Delta_{\II{1}{n}}+W\1_{\II{1}{y}}$ by $\delta_y$. By computing the determinant from such first column we see that
\begin{align*}
\det\left([-\Delta_{\II{1}{n}}+W\1_{\II{1}{y}}]_{1\to\delta_y}\right)&=(-1)^{y+1}\det\left( 
\begin{array}{c|c} 
  T & 0 \\ 
  \hline 
  M & -\Delta_{\II{1}{n-y}}
\end{array}\right)\\
&=(-1)^{y+1}\det(T)\det(-\Delta_{\II{1}{n-y}})=n-y+1,
\end{align*}
since $T$ is a lower triangular square matrix of size $y-1$ with $(-1)$ on all the diagonal, and $\det(-\Delta_{\II{1}{k}})=k+1$ for all $k\in\N$ (we use the convention $\det(-\Delta_{\emptyset})=1$).

Consider $\det(-\Delta_{\II{1}{n}}+W\1_{\II{1}{y}})$ as a polynomial in $(W(j))_{j=1}^y$. It is clear that it does not contain squares, or grater powers, of any $W(j)$. Moreover, the coefficient of the monomial $W(j_1)W(j_2)\cdots W(j_{k-1})W(j_k)$ with $1\leq j_1<j_2<\ldots j_{k-1}<j_k\leq y$ and $k\geq1$ can easily be computed as
\begin{align*}
    \det(-\Delta_{\II{1}{j_1-1}})&\det(-\Delta_{\II{j_1+1}{j_2-1}})\cdots\det(-\Delta_{\II{j_{k-1}+1}{j_k-1}})\det(-\Delta_{\II{j_k+1}{n}})\\
    &=j_1(j_2-j_1)\cdots(j_k-j_{k-1})(n+1-j_k).
\end{align*}
The remaining constant coefficient is $\det(-\Delta_{\II{1}{n}})=n+1$, which means all coefficients of $\det(-\Delta_{\II{1}{n}}+W\1_{\II{1}{y}})$ are positive and therefore
\begin{align*}
    G_{\II{1}{n},W}(1,y)&\leq\frac{n+1-y}{\det(-\Delta_{\II{1}{n}}+W\1_{\II{1}{y}})}\leq\frac{n+1-y}{\sum_{j=1}^y j(n+1-j)W(x)}\\
    &\leq\left(\sum_{j=1}^y j W(j)\right)^{-1}=\left(\sum_{j=0}^{y-1} (y-j) W(y-j)\right)^{-1}.\qedhere
\end{align*}
\end{proof}

With the previous proposition in mind we define for $\delta>0$ and $x\in\Z^d$ 
\begin{align*}
    Z^+_\delta(x)&\coloneqq\min\left\{n\in\N\,\middle|\,\sum_{j=1}^n(n+1-j)V(x+j)>\delta^{-1}\right\},\\
    Z^-_\delta(x)&\coloneqq\min\left\{n\in\N\,\middle|\,\sum_{j=1}^n(n+1-j)V(x-j)>\delta^{-1}\right\},\\
    A_\delta(x)&\coloneqq\II{x-Z^-_\delta(x)}{x+Z^+_\delta(x)}.
\end{align*}
Notice that $V(x)$ is not included in the definition of $Z^\pm_\delta(x)$ and therefore $Z^+_\delta(x)$ and $Z^-_\delta(x)$ are independent for all $x\in\Z$. 

It follows from \eqref{EqGRI}, the definitions above, potential monotonicity, and Propositions \ref{PrEVLF}, \ref{PrGF} that
\begin{align*}
    \lambda_{n,V}\norm{L_{n,V}}_\infty&\leq \lambda_{n,V}\max_{x\in\Lambda_n}\left[L_{A_\delta(x),0}(x)+2\delta\norm{L_{n,V}}_\infty\right]\\
    &\leq \lambda_{n,V}\max_{x\in\Lambda_n}\norm{L_{A_\delta(x),0}}_\infty+2\delta C(1).
\end{align*}
By domain monotonicity and translation invariance, the last maximum above is attained at the $x\in\Lambda_n$ that also maximises $\#A_\delta(x)=Z^+_\delta(x)+Z^-_\delta(x)+1$. Moreover, $V$ being i.i.d. implies $\lim_{n\to\infty}\max_{x\in\Lambda_n}Z^+_\delta(x)+Z^-_\delta(x)=\infty\quad\mathbb{P}$-a.s. and therefore Proposition \ref{PrLFB} and Theorem \ref{Th1} give
\begin{equation*}
    \varlimsup_{n\to\infty}\lambda_{n,V}\norm{L_{n,V}}_\infty\leq\frac{\mu_1}{2} \varlimsup_{n\to\infty}\left(\frac{\max_{x\in\Lambda_n}Z^+_\delta(x)+Z^-_\delta(x)}{2y_n}\right)^2+2\delta C(1)\quad\mathbb{P}\text{-a.s.}
\end{equation*}
The proof of Theorem \ref{Th2} ii) is finished with the next proposition followed by the limit $\delta\to0$.

\begin{Proposition}
For all $\delta>0$, $\displaystyle \varlimsup_{n\to\infty}\frac{1}{2y_n}\max_{x\in\Lambda_n}[Z^+_\delta(x)+Z^-_\delta(x)]\leq1\quad\mathbb{P}$-a.s.
\end{Proposition}
\begin{proof}
We will prove this over an exponential subsequence; the extension is done as in the proof of Proposition \ref{PrY} using the monotonicity of $n\mapsto\max_{x\in\Lambda_n}[Z^+_\delta(x)+Z^-_\delta(x)]$. In addition to $Z^+_\delta(x)$ being independent of $Z^-_\delta(x)$ for all $x\in\Z^d$, we also have that all $Z^\pm_\delta(x)$ are equal in distribution to $Z^+_\delta(0)$.

Assume \textbf{(C1)}. For all $t>0$ we have
\begin{equation*}
    \EE{e^{-t V(0)}}=\EE{e^{-t V(0)}\1_{V(0)\leq \frac{1}{\sqrt{t}}}}+\EE{e^{-t V(0)}\1_{V(0)> \frac{1}{\sqrt{t}}}}\leq F\left(1/\sqrt{t}\right)+e^{-\sqrt{t}}.
\end{equation*}
With this, we use the exponential Markov inequality and independence to obtain
\begin{align*}
    \PPrb{Z_\delta^+(0)>n}&=\PPrb{\sum_{j=1}^n(n+1-j)V(x+j)^{-1}\leq\delta^{-1}}\\
    &\leq e^{t/\delta}\prod_{j=1}^n\EE{\exp(-t j V(0))}\leq e^{t/\delta}\left(F\left(1/\sqrt{t}\right)+e^{-\sqrt{t}}\right)^n,\qquad n \in\N.
\end{align*}
Now we proceed with the distribution of $Z^+_\delta(0)+Z^-_\delta(0)$ as
\begin{align*}
    \PPrb{Z^+_\delta(0)+Z^-_\delta(0)>n}&=\PPrb{Z^+_\delta(0)>n-1}+\sum_{j=1}^{n-1}\PPrb{Z^+_\delta(0)=j}\PPrb{Z^-_\delta(0)>n-j}\\
    &\leq 2\PPrb{Z^+_\delta(0)>n-1}+\sum_{j=2}^{n-1}\PPrb{Z^+_\delta(0)>j-1}\PPrb{Z^+_\delta(0)>n-j}\\
    &\leq 2e^{t/\delta}\left(F\left(1/\sqrt{t}\right)+e^{-\sqrt{t}}\right)^{n-1}+(n-2)e^{2t/\delta}\left(F\left(1/\sqrt{t}\right)+e^{-\sqrt{t}}\right)^{n-1}\\
    &\leq n e^{2t/\delta}\left(F\left(1/\sqrt{t}\right)+e^{-\sqrt{t}}\right)^{n-1}.
\end{align*}
For any $\epsilon>0$ define $t(\epsilon)$ by $\ln\left(F\left(1/\sqrt{t(\epsilon)}\right)+e^{-\sqrt{t(\epsilon)}}\right)\leq \frac{\ln F(0)}{1+\epsilon}$ so that
\begin{align*}
    \PPrb{\max_{x\in\Lambda_n}[Z^+_\delta(x)+Z^-_\delta(x)]>\floor{(1+\epsilon)^2 2y_n}}&\leq C(1)n\,\PPrb{Z^+_\delta(0)+Z^-_\delta(0)>\floor{(1+\epsilon)^2 2y_n}}\\
    &\leq C(1) e^{2t(\epsilon)/\delta} n \exp\left[(1+\epsilon)^2 2 y_n\frac{\ln F(0)}{1+\epsilon}(1+o(1))\right]\\
    &=C(1)e^{2t(\epsilon)/\delta}n^{-\epsilon(1+o(1))},
\end{align*}
which is summable over the exponential subsequence $n=\floor{e^m}$, $m\in\N$.

Assume \textbf{(C2)}. We follow the same steps as for \textbf{(C1)}  above. To bound the Laplace transform of $V(0)$ we consider the function $f(t)\coloneqq a[F(t)]^{1/\eta}$ for some $a>0$. From \textbf{(C2)} follows that there exists $t_0\in(0,\infty)$ such that $F(t)\leq 2\,c\,t^\eta$ for all $t\in[0,t_0]$. Therefore, by choosing $a\coloneqq(t_0^{-1}+(2c)^{1/\eta})^{-1}$ we obtain
\begin{equation*}
    0\leq f(t)\leq\begin{cases}
    a(2c)^{1/\eta}t\leq t,&t\in[0,t_0],\\
    a\leq t_0\leq t, &t\in(t_0,\infty).
    \end{cases}
\end{equation*}
Moreover, since $\PPrb{f(V(0))\leq t}=\left(\frac{t}{a}\right)^\eta$ for $t\in[0,a]$, we have
\begin{align*}
        \EE{\exp(-t V(0))}&\leq\EE{\exp(-t f(V(0)))}=\frac{\eta}{(a)^{\eta}}\int_0^{a} e^{-t y} y^{\eta-1}\dd{y}\\&\leq\frac{\eta}{(a)^{\eta}}\int_0^{\infty} e^{-t y} y^{\eta-1}\dd{y}= \frac{\eta\Gamma(\eta)}{(a t)^\eta},\qquad t>0.
\end{align*}
The exponential Markov inequality at $t=n\eta\delta$, independence, and the Stirling bound $(n/e)^n\leq n!$ lead us to
\begin{align*}
    \PPrb{Z_\delta^+(0)>n}&\leq e^{t/\delta}\prod_{j=1}^n\EE{\exp(-t j V(0))}\leq \left(\frac{\eta\Gamma(\eta)}{(at)^\eta}\right)^n\frac{e^{t/\delta}}{(n!)^{\eta}}\\
    &\leq \left(\frac{\eta^{1-\eta}\Gamma(\eta)e^{2\eta}}{a^\eta\delta}\right)^n n^{-2\eta n}\eqqcolon K_\delta^n \,n^{-2\eta n},\qquad n \in\N,
\end{align*}
from which follows
\begin{equation*}
    \PPrb{Z^+_\delta(0)+Z^-_\delta(0)>n}
    \leq2 K_\delta^{n-1}(n-1)^{-2\eta(n-1)}+K_\delta^{n-1} \sum_{j=2}^{n-1} \left(j-1\right)^{-2\eta(j-1)}(n-j)^{-2\eta(n-j)}.
\end{equation*}
The function $[2,n-1]\ni j\mapsto\left(j-1\right)^{-(j-1)}(n-j)^{-(n-j)}$ attains its unique maximum at $j=(n+1)/2$, therefore
\begin{align*}
    \PPrb{Z^+_\delta(0)+Z^-_\delta(0)>n}&\leq2 K_\delta^{n-1}(n-1)^{-2\eta(n-1)}+(4^{\eta}K_\delta)^{n-1}(n-2)(n-1)^{-2\eta(n-1)}\\
    &\leq (4^{\eta}K_\delta)^{n-1}n(n-1)^{-2\eta(n-1)}.
\end{align*}
Finally, for $\epsilon>0$ we have
\begin{align*}
    \PPrb{\max_{x\in\Lambda_n}[Z^+_\delta(x)+Z^-_\delta(x)]>\floor{(1+\epsilon)2y_n}}&\leq C(1)n\,\PPrb{Z^+_\delta(0)+Z^-_\delta(0)>\floor{(1+\epsilon)2y_n}}\\
    &= C(1) n \exp\left[-(1+\epsilon)4\eta y_n(\ln y_n)(1+o(1))\right]\\
    &= C(1) n \exp\left[-(1+\epsilon)4\eta y_n(\ln \ln n)(1+o(1))\right]\\&=C(1)n^{-\epsilon(1+o(1))},
\end{align*}
which is summable over the exponential subsequence $n=\floor{e^m}$, $m\in\N$.
\end{proof}

\appendix

\section{Proof of Theorem \ref{ThSG}}
Let $(X_t)_{t\geq0}$ be a continuous time simple symmetric random walk on $\Z^d$ with jump intensity $1$, and let $\mathrm{P}_x$, $\mathrm{E}_x$ be the associated probability measure and expectation conditioned on $X_0=x$. We remark that $(X_t)_{t\geq0}$ is the Markov process that generates $-\Delta/(2d)$ on $\ell^2(\Z^d)$.

For a finite $A\subseteq\Z^d$ and $W:A\rightarrow[0,\infty)$, the Feynman–Kac formula lets us write the semigroup generated by $-\Delta_{A}+W$ as 
\begin{equation*}
    \exp\left(-\frac{t}{2d}[-\Delta_{A}+W]\right)\phi(x)=\E{x}{\phi(X_{t})\exp\left(-\int_0^{t}\frac{W(X_s)}{2d}\dd{s}\right)\1_{t<\tau_A}},\quad\phi\in\ell^2(A),\,x\in A,
\end{equation*}
where $\tau_A\coloneqq\inf\{t\geq0\,|\,X_t\notin A\}$ is the exit time of $A$. To simplify notation we set $\lambda=\frac{\lambda_{A,W}}{2d}$, $K_t=\exp\left(-\frac{t}{2d}[-\Delta_{A}+W]\right)$ and $k_t(x,y)=\hp{\delta_x}{K_t \delta_y}_{\ell^2(A)}$ (the kernel of the semigroup). Depending on  $\lambda$ we distinguish two cases.
\begin{itemize}[leftmargin=*]
\item Case $\lambda\leq\frac{1}{d}$.
\end{itemize}
Let $\cc{B_\infty}(x,r)\coloneqq\{y\in\R^d\,|\,\norm{x-y}_\infty\leq r\}$. For $t\geq 1$ and $x\in A$ we have
\begin{align*}
K_{t/\lambda}\1_A(x)&=K_{t/\lambda}\1_{A\cap \cc{B_\infty}(x,r)}(x)+K_{t/\lambda}\1_{A\setminus\cc{B_\infty}(x,r)}(x)\\
&\leq\hp{k_{1/\lambda}(x,\cdot)}{K_{(t-1)/\lambda}\1_{A\cap \cc{B_\infty}(x,r)}}_{\ell^2(A)}+\Prb{0}{X_{t/\lambda}\notin\cc{B_\infty}(0,r)}\\
&\leq C(d)\sqrt{\Prb{0}{X_{2/\lambda}=0}}r^{d/2}e^{-(t-1)}+\Prb{0}{X_{t/\lambda}\notin\cc{B_\infty}(0,r)},
\end{align*}
where we chose $r=2t\sqrt{\frac{e}{\lambda d}}$.

The term $\Prb{0}{X_{2/\lambda}=0}$ can be estimated using the characteristic function of $X_s$, which is $\phi_s(\theta)=\exp\left(-s+\frac{s}{d}\sum_{i=1}^d\cos(\theta_i)\right)$, by means of
\begin{equation*}
    \Prb{0}{X_{s}=0}=\frac{1}{(2\pi)^d}\int_{[-\pi,\pi]^d}\phi_s(\theta)\dd{\theta}=\frac{e^{-s}}{(2\pi)^d}\left[\int_{[-\pi,\pi]}\exp\left(\frac{s}{d}\cos\theta_1\right)\dd{\theta_1}\right]^d.
\end{equation*}
Laplace's method applied to the right-most integral yields $\Prb{0}{X_{s}=0}s^{d/2}\xrightarrow[s\to\infty]{}\left(\frac{d}{2\pi}\right)^{d/2}$ and therefore
$\Prb{0}{X_{2/\lambda}=0}\leq C(d)\lambda^{d/2}$.

For the other probability we use the bound (see \cite[Lemma 4.6]{Barlow})
\begin{equation*}
    P[S_n>y]\leq e^{-y^2/(2n)},\qquad y>0,
\end{equation*}
where $S_n$ is a discrete time simple symmetric random walk on $\Z$ starting at $0$, and $P$ is its probability measure. Recalling that the first component of $X_{t}$, which we denote $X_t^1$, is a continuous time simple symmetric random walk on $\Z$ with jump intensity $1/d$ we have
\begin{equation*}
    \Prb{0}{X_{t/\lambda}\notin \cc{B_\infty}(0,r)}\leq 2d \Prb{0}{X_{t/\lambda}^1>r}\leq 2d\sum_{n\geq 0}\frac{e^{-t/(\lambda d)}}{n!}\left(\frac{t}{\lambda d}\right)^n P[S_n>r].
\end{equation*}
We split the series at $n=\frac{2e t}{\lambda d}$ and bound the two terms separately:
\begin{align*}
    \sum_{n\leq \frac{2e t}{\lambda d}}\frac{e^{-t/(\lambda d)}}{n!}\left(\frac{t}{\lambda d}\right)^n P[S_n>r]&\leq \sum_{n\leq \frac{2e t}{\lambda d}}\frac{e^{-t/(\lambda d)}}{n!}\left(\frac{t}{\lambda d}\right)^n e^{-r^2/(2n)}\\&\leq \exp\left(-\frac{r^2\lambda d}{4 e t}\right)\sum_{n\leq \frac{2e t}{\lambda d}}\frac{e^{-t/(\lambda d)}}{n!}\left(\frac{t}{\lambda d}\right)^n \leq \exp\left(-\frac{r^2\lambda d}{4 e t}\right)=e^t,\\
    \sum_{n\geq\frac{2e t}{\lambda d}}\frac{e^{-t/(\lambda d)}}{n!}\left(\frac{t}{\lambda d}\right)^n P[S_n>r]&\leq e^{-t/(\lambda d)}\sum_{n\geq\frac{2e t}{\lambda d}}\frac{1}{n!}\left(\frac{t}{\lambda d}\right)^n\\
    &\leq e^{-t/(\lambda d)}\sum_{n\geq 0}\frac{1}{n!}\left(\frac{n}{2 e}\right)^n\leq C(1) e^{-t}.
\end{align*}

With this we have shown $\norm{K_{t/\lambda}\1_A}_\infty\leq C(d)\left(1+t^{d/2}\right)e^{-t}$ for $t\geq1$. Since $K_{t/\lambda}\1(x)$ is always bounded by $1$ we can add $\left(\inf_{0\leq t\leq1}\left(1+t^{d/2}\right)e^{-t}\right)^{-1}$ to $C(d)$, if necessary,  to have
\begin{equation*}
    \norm{K_{t/\lambda}\1_A}_\infty\leq C(d)\left(1+t^{d/2}\right)e^{-t},\qquad t\geq0.
\end{equation*}
Replacing $t$ by $2d\lambda t$ and gives the desired bound.
\begin{itemize}[leftmargin=*]
\item Case $\lambda\geq\frac{1}{d}$.
\end{itemize}
This case follows form the heat kernel bound (see \cite[Theorem 5.17]{Barlow})
\begin{equation*}
    \Prb{0}{X_t=y}\leq C(d) \exp\left[-t-\norm{y}_1\ln\left(\frac{\norm{y}_1}{e t}\right)\right],\qquad \norm{y}_1\geq e t.
\end{equation*}
We proceed as before but now we use $\cc{B_1}(x,r)\coloneqq\{y\in\R^d\,|\,\norm{x-y}_1\leq r\}$. For $t\geq 0$ we have
\begin{align*}
K_{t}\1_A(x)&=K_{t}\1_{A\cap \cc{B_1}(x,r)}(x)+K_{t}\1_{A\setminus\cc{B_1}(x,r)}(x)\\
&\leq\hp{\delta_x}{K_{t}\1_{A\cap \cc{B_1}(x,r)}}_{\ell^2(A)}+\Prb{0}{X_t\notin\cc{B_1}(0,r)}\\
&\leq C(d) r^{d/2}e^{-\lambda t}+\Prb{0}{X_t\notin\cc{B_1}(0,r)},
\end{align*}
with $r=\lambda t d e^2$. Clearly $r\geq e t$, so we can apply the heat kernel bound to obtain
\begin{align*}
    \Prb{0}{X_{t}^1\notin \cc{B_1}(0,r)}&\leq C(d)\sum_{\substack{y\in\Z^d\\\norm{y}_1>r}}\exp\left[-t-\norm{y}_1\ln\left(\frac{\norm{y}_1}{e t}\right)\right]\\&\leq C(d)\sum_{\substack{y\in\Z^d\\\norm{y}_1>r}}\exp\left[-\norm{y}_1\ln(e d)\right]\\&\leq C(d) e ^{-r}\sum_{y\in\Z^d}\exp\left[-\norm{y}_1\right]\\
    &= C(d) e^{-r}\leq C(d) e^{-\lambda t},
\end{align*}
and therefore $\norm{K_{t}\1_A}_\infty\leq C(d)\left(1+[\lambda t]^{d/2}\right)e^{-\lambda t}$. Replacing $t$ by $2d t$ and gives the desired bound.

\newcommand{\etalchar}[1]{$^{#1}$}

\end{document}